\documentclass[12pt]{iopart}
\usepackage{graphics,epsfig,colordvi}
\def\be{\begin{eqnarray}}
\def\ee{\end{eqnarray}}

\def\dsp{\displaystyle}

\def\lsim{\stackrel{\scriptstyle <}{\phantom{}_{\sim}}}
\def\gsim{\stackrel{\scriptstyle >}{\phantom{}_{\sim}}}
\begin{document}
\title{Resonances from a Hadronic Fireball}
\author{Evgeni E. Kolomeitsev\dag and Peter Filip\ddag}
\address{\dag\ European Centre for Theoretical Studies in Nuclear
Physics and Related Areas\\ Villa Tambosi, I-38050 Villazzano (TN)
and INFN, G.C. Trento , Italy}
\address{\ddag\ Max-Planck-Institut f\"ur Physik, D-80805  Munich,
Germany}

\begin{abstract}
Production of $\phi(1020), \Lambda^*(1520)$ and $\overline{K^*}(892)$ resonances
at the final stage of a heavy-ion collision is considered.
It is shown that original momentum distributions and abundance of resonances
formed during the process of heavy ion collision may differ significantly
from their measured spectra and yields. Reconstruction probability of resonances
decaying inside the fireball can be strongly  suppressed because of
interactions of their hadronic decay products in the fireball medium.
We investigate dependence of the degree of the suppression on
the fireball size, dynamics, and the resonance decay width in medium.
Quantitative results are presented for lead-lead collisions
at 158~GeV SPS beam energy.
\end{abstract}
\section{Introduction}

Present experimental information available in the field of
heavy-ion collisions allows for the systematic investigation of
strongly interacting hadronic matter under extreme conditions.
High statistics accumulated in various experiments facilitates
reconstruction of hadronic resonances via their decay products. At
CERN-SPS experiments several resonances have been identified so
far: $\phi(1020)$ mesons are detected by NA49 collaboration via
$\phi\to K^+ K^-$ decay ~\cite{na49} and by NA50 via $\phi\to
\mu^+\mu^-$ channel ~\cite{na50}, $\Lambda^*(1520)$ and
$\overline{K^*}(892)$ are seen by NA49 in their dominant decay
modes~\cite{markert,sammer,friese}, the $\rho$ and $\omega$ mesons
are measured by NA50 in dimuon channel~\cite{quintas}. Already the
first preliminary data were surprising. Total averaged $\phi$
meson multiplicity extracted from NA49 data was found to be
significantly smaller than that obtained from the extrapolation of
preliminary NA50 data. Another interesting observation is that
preliminary NA49 data on $\Lambda^*(1520)$ multiplicity reveal
significant suppression of the yield per participating nucleon
when compared to inelastic proton-proton collisions. This
contradicts to systematics of other strange particle abundance at
SPS energies.

Suggested explanation of the $\phi$-meson puzzle~\cite{shur99} was
that $\phi$-mesons decaying inside a
fireball can disappear from the $K^+K^-$ mass peak due to rescattering
and absorption  of secondary kaons in the surrounding medium.
Such mechanism has been quantitatively studied in~\cite{suny} where
suppression at the level 40--60\% has been obtained from simulation
using RQMD code. Similar mechanism could also account for
$\Lambda^*(1520)$ suppression.
In \cite{soff} 75\% suppression for $\phi(1020)$ and
50\% for $\Lambda^*(1520)$ particles have been obtained within
UrQMD model. Therefore the rescattering of resonance decay products
can indeed explain in part some experimental observations.
However, besides  the lack of quantitative account of the
difference in the extrapolated $\phi$ meson yields,
other caveat has been issued in~\cite{tr01}:
Since the vacuum width of $\overline{K^*}(892)$ mesons is more than 3 times
larger than the $\Lambda^*(1520)$ width, $\overline{K^*}(892)$ mesons
decay inside a fireball more frequently and
should be therefore  suppressed even stronger than $\Lambda^*(1520)$.
The preliminary data of NA49 \cite{sammer} do not show such a
suppression for $\overline{K^*}(896)$.

In this paper we discuss to what extent the in-medium modifications of the
resonance decay width can effect the resonance suppression and
help to accommodate empirical information.

\section{Global picture of a collision}
In our study we consider a two-stage picture of the hadronic phase of
heavy-ion collision at SPS energies which can be
characterized by the notions of chemical and thermal
freeze-outs~\cite{pbm,cr,heinz}. At the initial stage of
the hadronic phase we assume the temperature $T$
to be close to the QCD  phase transition temperature
$T_c\approx 170\pm 10$~MeV. Then
the system expands up to the point, when numbers of
different kinds of particles freeze in a chemical
freeze-out. Thermodynamical parameters of this stage could be
obtained by fitting the final total hadron multiplicities. Typical
temperature is found to be $T_{\rm chem}\sim 160\pm 10$~MeV~\cite{pbm,cr}.
During the second stage of collision, elastic scattering
processes change momentum distributions of hadrons in accord with a
decreasing  temperature until they cease and
distributions freeze in. The freeze-out temperature can be
extracted from a simultaneous fit to the single-particle
$m_T$-spectra of different particles supplemented by the analysis
of particle correlation data.
Typical temperatures are found to be $T=T_{\rm therm}\sim
110\pm 30$~MeV~\cite{heinz}.

The key assumption behind the decay product rescattering (DPR) mechanism
of \cite{shur99} is that the final resonance distribution
is formed at some stage between the chemical and thermal
freeze-out. At this moment, mean free paths of
the resonance, $\lambda_r$, pions, kaons and nucleons,
$\lambda_{\pi, K,N}$, and the characteristic size of the fireball, $R$,
should  satisfy
\be\label{mfp}
\lambda_{\pi},\lambda_{K},\lambda_N\ll R \lsim\lambda_r\,.
\ee
This implies that
upon this stage resonances stream freely out from the fireball.
For their decay products, pions and kaons, on the other hand,
the surrounding medium is still opaque.
Therefore, if a resonance decays inside a fireball in hadronic
channel, the decay products  can be rescattered or absorbed, so
that the resonance remains unobserved experimentally.
It is now obvious that the possible increase of a resonance decay width
in medium will enlarge the probability of its decay inside a
fireball enhancing thereby the DPR mechanism.
Besides the total decay width of the resonance
and the typical size of the fireball, one of the crucial parameters
controlling the efficiency of the DPR mechanism is,  the life time of the fireball
after the resonance freeze-out, $\tau_{\rm f.o.}$.

\section{Apparent resonance distribution}

We denote the phase-space distribution of  the resonance $(r)$ in the
center-of-mass system of two colliding nuclei at the moment of its thermal
freeze-out as
$f^{(r)}(\vec{x},\vec{p})$. Then the primary momentum distribution
of the resonance is
\be\label{INTEGR} \eta_0^{(r)}(p) = \langle\, 1\,\rangle
\,,\quad
\Big\langle \dots \Big\rangle =
\intop_{\Sigma}
{\rm d}^3 {\sigma}^\mu\,p_\mu\, f^{(r)}(\vec{x},\vec{p\,})
\Big(\dots\Big)\,.
\ee
where integration goes over
the fireball volume within a freeze-out hyper-surface $\Sigma$.
In the absence of any in-medium modification of the resonance and
rescattering of its decay products,
the shape of the  momentum
distribution {\it observed} in the decay channel $j$ is given by
$\eta_0^{(r)}(p)\,\Gamma^{(r)0}_{j}/\Gamma^{(r) 0}_{\rm tot}$,
where $\Gamma^{(r) 0}_j$ and $\Gamma^{(r) 0}_{\rm tot}$
are the partial and total decay widths of the resonance,
respectively.
Accordingly, in the experimental analysis, the resonance
distribution would be reconstructed from momentum distribution of decay
products multiplied by the corresponding inverse branching ratio.

Taking into account modifications of partial and total widths of
the resonance and the rescattering of decay products in medium we
write expression for the observed resonance distributions in the
decay channel $j$ in the following form
\be  \eta_j^{(r)}(p)
&=&\Big\langle{\rm D}(\tau)
+\frac{\widetilde{\Gamma}_{\rm tot}^{(r) 0}} {\widetilde{\Gamma}_{j}^{(r) 0}}
\, \intop_0^{\dsp \tau}\! {\rm d}t \,{\rm D}(t)\,
\widetilde{\Gamma}_{j}^{(r) *}(t) \, {\rm P}^{(r)}_{j \lambda}(t)\, {\rm
P}_{\rm rec} \Big\rangle \,.
 \label{ratrescat}
\ee
Here
\be\label{PBX}
{\rm D}(t)&=&\exp\Big[-\int_0^{t}
\widetilde{\Gamma}_{\rm tot}^{(r) *}(t')\,
{\rm d} t' \Big]\,,
\ee
is the probability that resonance will fly for a time $t$
starting from a position $\vec{x}$ where it suffers the last
interaction.
$\widetilde{\Gamma}^{(r)}_{\rm tot}=\Gamma^{(r)}_{\rm tot}\,m_r/E_r$
is the total width of a moving particles with energy
$E_r=(m_r^2+p^2)^{1/2}$, where $m_r$ is a resonance mass. Here and
below asterisk  denotes in-medium values of the quantities, which
is determined  by the current local temperature and density.
The probability of the decay products to leave fireball
without any rescattering is ${\rm P}_{j \lambda}^{(r)}$. For the explicit
form of ${\rm P}_{j \lambda}^{(r)}$ we refer to Ref.~\cite{fk01}, where
it is written for the case of the two-particle decay. The quantity
${\rm P}_{\rm rec}$ stands for the probability to identify a
resonance from non-rescattered decay product. Without in-medium
effects we put $P_{\rm rec}=1$ assuming an ideal detector. If
spectra of resonance decay products differ in medium from the
vacuum ones, the momenta of secondary particles will change on the
way out from the fireball even without being rescattered, since
in this case a fireball serves as a potential well.
This effect can be especially strong for strange resonances since
the properties of kaons in the final state are strongly modified
in nuclear matter or/and isospin asymmetrical pion
gas~\cite{kmed,lk,lkk}. For a rough estimation of the maximal suppression
effect we put in this case ${\rm P}_{\rm rec}=0$.
The time scale $\tau$ in (\ref{ratrescat}) is the time spent by the
resonance in medium. It is given by
$\tau=\min\{\tau_{r}(\vec{v},\vec{x}),\tau_{\rm f.o.}\}$, where
$\tau_r(\vec v,\vec x)$ is the time during which the resonance
with a velocity $\vec v$ flies from the position $\vec{x}$ till the
border of the fireball.
Finally, in (\ref{ratrescat}) we integrate  over all initial resonance
positions $\vec x$ with the phase space distribution
(\ref{INTEGR}).

In the case when the partial width of decay channel $j$ does not change in medium
$\Gamma_j^{(r)*}=\Gamma_j^{(r)0}$, decay products are not rescattered
${\rm P}_{j\lambda}^{(r)}=
{\rm P}_{\rm rec}=1$ and the total width $\Gamma_{\rm tot}^{(r)*}
\neq \Gamma_{\rm tot}^{(r)0}$, we have
\be \label{ratmu}
\eta_j^{(r)}(p) &=&
\Big\langle{\rm D}(\tau)+\widetilde{\Gamma}_{\rm
tot}^{(r) 0}\,
 \intop_0^{\dsp \tau}
{\rm d}t\, {\rm D}(t)
\Big\rangle\,.
\ee
This expression corresponds, e.g., to the case when resonances are
reconstructed in the leptonic channels.
Note that for $\Gamma_{\rm tot}^{(r) *}=\Gamma_{\rm tot}^{(r) 0}$ we have
$\eta_\mu^{(r)}\equiv \eta_0^{(r)}$\,, for $\Gamma_{\rm
tot}^{(r) *}>\Gamma_{\rm tot}^{(r) 0}$ we have $\eta_\mu^{(r)}<\eta_0^{(r)}$
and if $\Gamma_{\rm tot}^{(r) *}<\Gamma_{\rm tot}^{(r) 0}$
then $\eta_\mu^{(r)} > \eta_0^{(r)}$.

After the general considerations we specify the model of the fireball expansion,
which is used in our numerical calculations. We assume a homogeneous spherical
fireball with the constant density and temperature profiles.
The primordial resonance momentum distribution
\be\label{dist}
f_r(\vec{x},\vec{p})=\exp\Big[-\frac{E_r
-\vec{p}\cdot \vec{u\,}(\vec{x\,})}
{T_0\,\sqrt{1-u^2(\vec{x\,})}}\Big]
\,,\ee
is determined by the temperature $T_0$, flow velocity profile
$\vec{u}(\vec{x})=v_f\,\vec{x}/R_0$ and radius $R_0$.
The fireball expands
$R(t)=R_0+v_f\,t$\,, its density evolves as
$\rho(t)=\rho_0\,R_0^3/R^3(t)$ and the temperature decreases as
$T(t)=T_0\,R_0/R(t)$ as expected for relativistic pion gas,
cf.~\cite{voskjetp}.
Time of flight of a resonance inside a fireball starting from a
position $\vec{x}$ till the border is given by
\be \nonumber
\tau_r(\vec{v},\vec{x}\,) &=&
\Big(
\sqrt{ \left(\vec{v\,}\cdot
\vec{x\,}-v_f\,R_0\right)^2+
\left(R^2_0-\vec{x\,}^2\right)\,
\left(\vec{v\,}^2-v_f^2\right)}
\\\label{tau}
&-&\left(\vec{v}\cdot
\vec{x\,}-v_f\,R_0\right)
\Big)\left(\vec{v\,}^2-v_f^2\right)^{-1}\,,
\ee
which  is valid for $|\vec{x}|<R_0$ and $|\vec{v}|>v_f$. In
the case $|\vec{v}|<v_f$ we put $\tau_r=\infty$.
Parameters $R_0$\,, $T_0$\,, $v_f$\,,
 serve as input for numerical evaluations below.

The particular realization of (\ref{ratrescat}) is
a rather crude approximation.
However, the final results are found \cite{fk01} to be rather
insensitive to the details of hydrodynamical evolution of a
fireball, being determined mainly by the values of $\Gamma_{\rm
tot}^{(r) *}\, R_0$, $\Gamma_{\rm tot}^{(r) *}\,
\tau_{\rm f.o.}$, and $v_f$.

\section{Applications}

First we discuss numerical results obtained in~\cite{fk01}
for $\phi $ meson yields reconstructed
via $K^+K^-$ and $\mu^+\mu^-$  channels in central Pb+Pb collisions
at 158~GeV/n SPS energy. Being dominantly an $s\bar s$ state,
the $\phi$ mesons can decouple  quite early, since  its interaction
with non-strange matter is suppressed according to OZI rule.
The mean free path $\lambda_\phi$ of $\phi$ mesons in hadron gas is
estimated in~\cite{haglinphi}. Comparison with mean free paths
of pions and kaons, $\lambda_{\pi,K}$, from~\cite{piklam} gives
$\lambda_\pi\lsim\lambda_K<\lambda_\phi$ for temperatures
$T_{\rm therm}<T<T_{\rm chem}$. Therefore condition (\ref{mfp})
can be satisfied.

For a comparison with experimental data we define the following suppression
factors
\be\label{etamt}\fl
\mathcal{R}(m_T)=\mathcal{R}_K(m_T)/\mathcal{R}_\mu(m_T)\,,\;\;
\mathcal{R}_{K,\mu}(m_T)=<\eta^{(\phi)}_{K,\mu}(p)>_y/<\eta^{(\phi)}_0(p)
>_y\,,
\ee
where $<\dots >_y$ means the integration over experimental rapidity
interval. The apparent distribution of $\phi$ mesons in the hadronic
channel ($K\bar K)$), $\eta^{(\phi)}_K$, is calculated according
to (\ref{ratrescat}). In the leptonic channel, $\eta_\mu^{(\phi)}$
we use (\ref{ratmu}).

First, we evaluate suppression factor (\ref{etamt})
without any modification of $\phi$ properties in medium.
In this case we have $\mathcal{R}_\mu(m_T)\equiv 1$ and
$\mathcal{R}(m_T)=\mathcal{R}_K(m_T)$.
We use several combinations of input parameters
$(T_0,R_0,v_f)$
chosen consistently with the $\phi$-meson freeze-out temperature
$T_0$ which we  vary between $T_{\rm chem}$ and $T_{\rm therm}$:
($i$) (150~MeV, 20~fm, 0.5) \,, ($ii$) (160~MeV, 15~fm, 0.46)\,,
($iii$) (170~MeV, 10~fm, 0.41).
Size of the fireball $R_0$ at the moment of $\phi$ freeze-out
has to be comparable with $\phi$ meson mean free path $\lambda _\phi
$. The above values for different temperatures are taken according to
estimations of $\lambda_\phi$ in~\cite{haglinphi}.
Flow velocities are  adjusted to reproduce the slope of $\phi $ meson $m_T$
distribution measured by the NA50 collaboration: $T_{\rm
eff}=218$~MeV. The fireball life time is determined by equation
$T(\tau_{\rm f.o.})=T_{\rm therm}$

\begin{figure}
\includegraphics[width=6cm,clip=true]{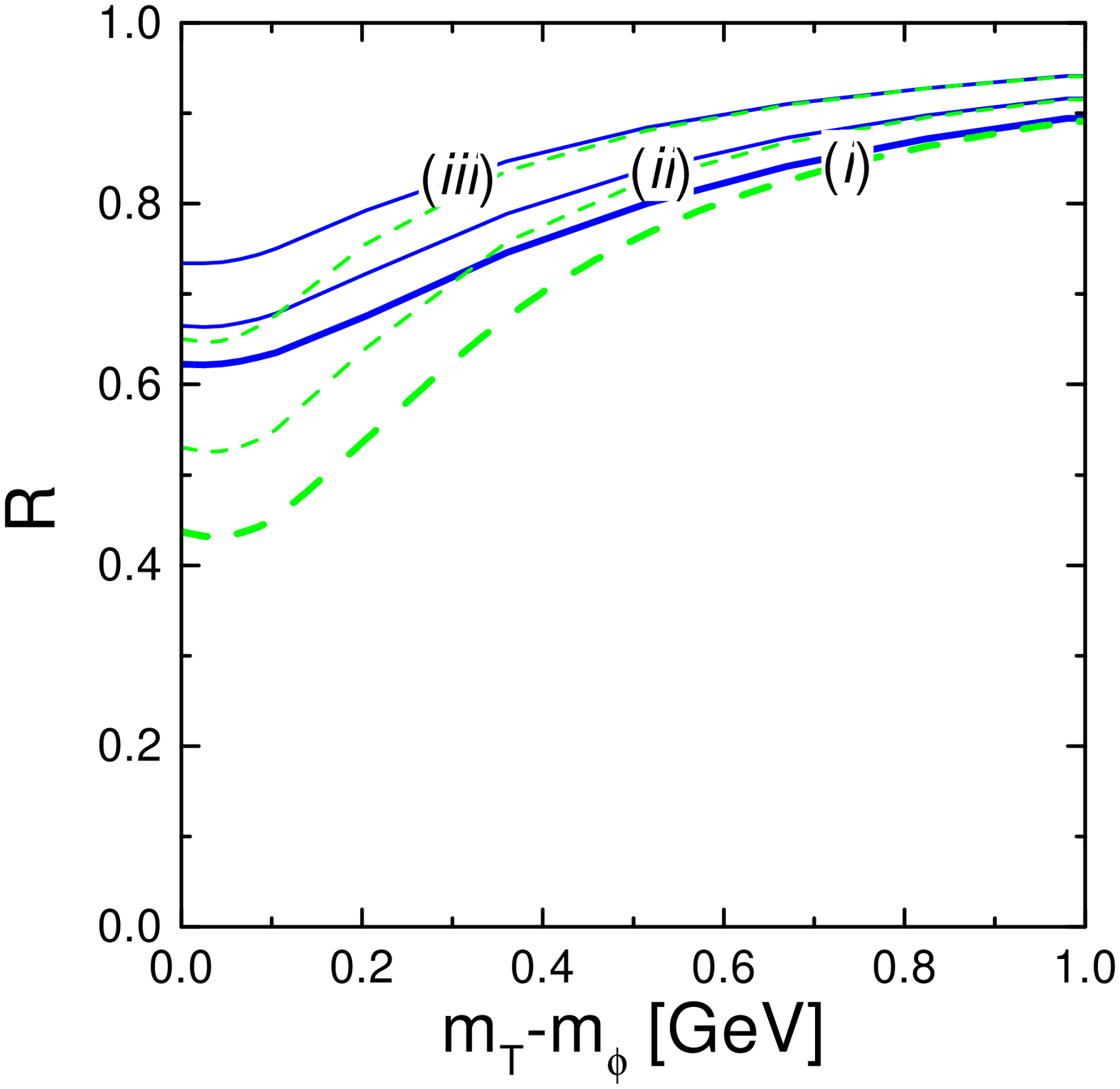}\qquad
\includegraphics[width=5.5cm,clip=true]{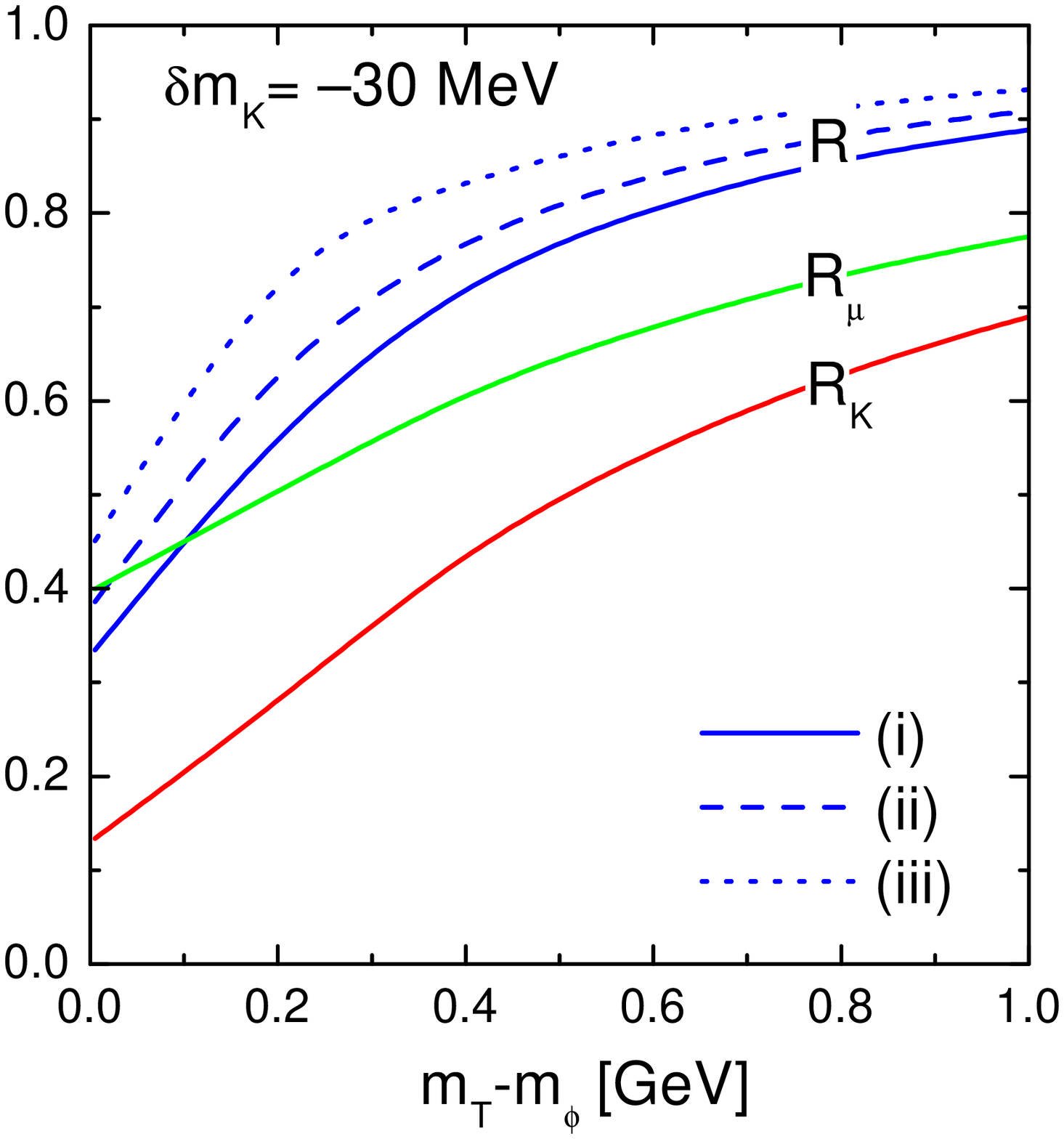}
\caption{{\it Left plane}: Ratio $\mathcal{R}$ calculated for
cases $(i)$--$(iii)$ without medium effects. Solid lines are
calculated with $T_{\rm therm}=80$~MeV and dash lines correspond
to $T_{\rm therm}=40$~MeV.\\ {\it Right plane}: Thick lines show
the ratio $\mathcal{R}(m_T)$ calculated with account for in-medium
modification of meson properties associated with a decreasing kaon mass
$\delta m_K^0=-30$~MeV Results for different parameters sets
$(i)$--$(iii)$ are depicted by solid, dash and dotted lines,
respectively. Thin lines illustrate the suppression factors in the
muon $\mathcal{R}_\mu$ and kaon $\mathcal{R}_K$ channels
calculated for parameter set ($i$).} \label{fig:phi}
\end{figure}

To reproduce results of RQMD calculations described in
Ref.~\cite{suny} we take freeze-out temperature
$T_{\rm therm}=80$~MeV, the kaon mean free path
$\lambda_K(t)=\lambda_K^0\,R_0^3/R^3(t)$ with
 $\lambda_K^0=0.5$~fm  and ${\rm P}_{\rm rec}=1$.
The temperature corresponds to the lowest
limit allowed by the analysis~\cite{heinz}. The results are shown in
Fig.~\ref{fig:phi} (left panel) by solid lines. The
limiting scenario considered in~\cite{suny}, when the freeze-out
volume is determined by the last kaon interactions, can be
reproduced with $T_{\rm therm}=40$~MeV. This case is shown by dash lines.
The solid lines in the left panel of Fig.~\ref{fig:phi} we take
as a reference point for our further investigation of  in-medium effects.

Let us now consider the case when the $\phi$ meson width increases
strongly in hadronic medium.
We simulate this effect by decreasing the kaon mass,
which can result e.g. from rescattering
of kaons on pions through $\overline{K^*}$ and heavier
kaonic resonances~\cite{shur}.
The modification of kaon properties in  medium
prevent the $\phi$ meson reconstruction even when kaons can leave
a fireball without a hard rescattering. Leaving the fireball,
kaons have to come back to their vacuum mass shell. Due to the energy conservation
the momenta of kaons change and, thereby, the invariant mass and
momentum of the pair. For this effect we account by putting ${\rm
P}_{\rm rec}=0$. Then the both suppression factors $\mathcal{R}_K$ and
$\mathcal{R}_\mu$ depend only  on the total $\phi$ meson width.
The $\phi$ total width reads  as $\Gamma_{\rm tot}^{(\phi)}(\delta m_K)=
\Gamma_{KK}\,p_{KK}^3(m_K+\delta m_K)/p_{KK}^3(m_K)+\Gamma_{\rho
\pi}$ with $p_{KK}(m_K)=\frac12\,m_\phi\,(1-4\,m_K^2/m_\phi^2)^{1/2}$.
The vacuum widths of $\phi\to K\bar K$ and $\phi\to \rho \pi$ decay
processes are equal to $\Gamma_{KK}=3.68$~MeV and $\Gamma_{\phi
\pi}=0.76$~MeV. We do not consider here in-medium change of the
$\Gamma_{\rho \pi}$ width, which effects weakly the resulting
suppression factor~\cite{fk01}.
We restrict ourselves to rather conservative modification
of kaon masses. Initially, at the $\phi$ freeze-out moment we
put $\delta m_K^0=-30$~MeV, and then it decreases linearly with the decreasing
fireball density $\delta m_K(t)=\delta m_K^0\, R_0^3/R(t)^3$.
It corresponds to the initial $\phi$ width $\Gamma_{\rm tot}^* \simeq
20$~MeV.
In this case $R_\mu<1$ and at small $m_T-m_\phi$ region
$R_\mu$ can be suppressed up to 40--60\%. Thus, for a given
freeze-out temperature $T_0$ we have to readjust the flow
velocity $v_f^0$ to reproduce the slope of the $m_T$ distribution
measured in dimuon channel by NA50 \cite{na50}.
For our three sets of parameters specified above
we obtain new flow velocities:
($i$) $v_f^0=0.38$\,, ($ii$)
$v_f^0=0.35$\,, ($iii$) $v_f^0=0.28$.

The results for a relative suppression of  hadronic and leptonic channels
are presented in the right plane of
Fig.~\ref{fig:phi}. Thick solid, dashed and dotted lines
drawn for cases ($i$)-($iii$) should be compared with the solid lines in the
left plane. We observe that the increase of the $\phi$ width
results in the overall increase of the suppression effect by about
20\%.
The suppression factors for leptonic and hadronic channels,
separately, are shown in Fig.~\ref{fig:phi} as well.
The increase of the $\phi$-meson width
provides a strong suppression of
$\mathcal{R}_K(m_T\to m_\phi)\sim 0.15$.
However, since the dimoun channel is also suppressed the resulting
ratio $\mathcal{R}$
remains on the level $\sim 0.3$ for small $m_T-m_\phi$.

\begin{figure}
\includegraphics[width=5.3cm,clip=true]{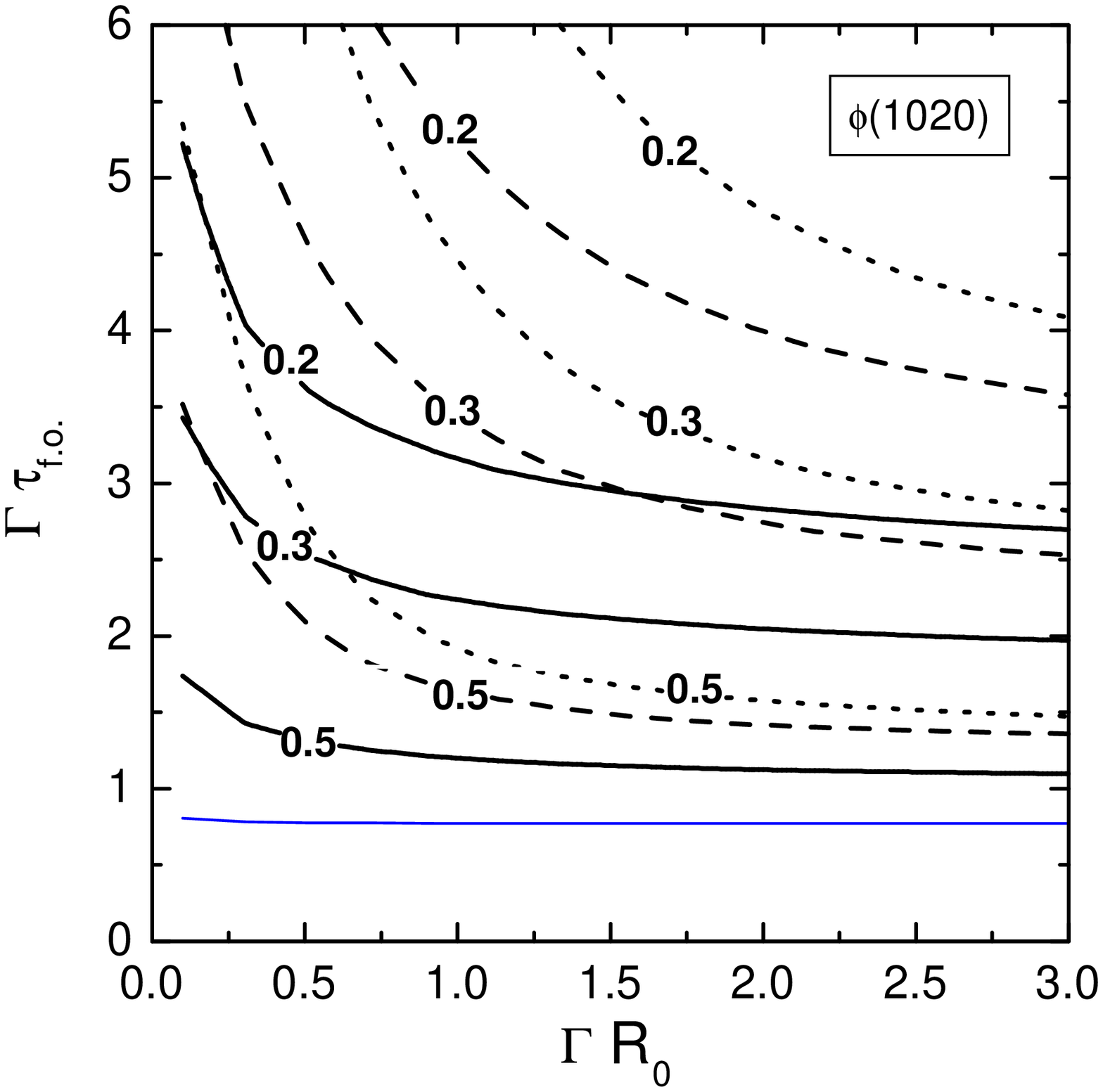}
\includegraphics[width=5.3cm,clip=true]{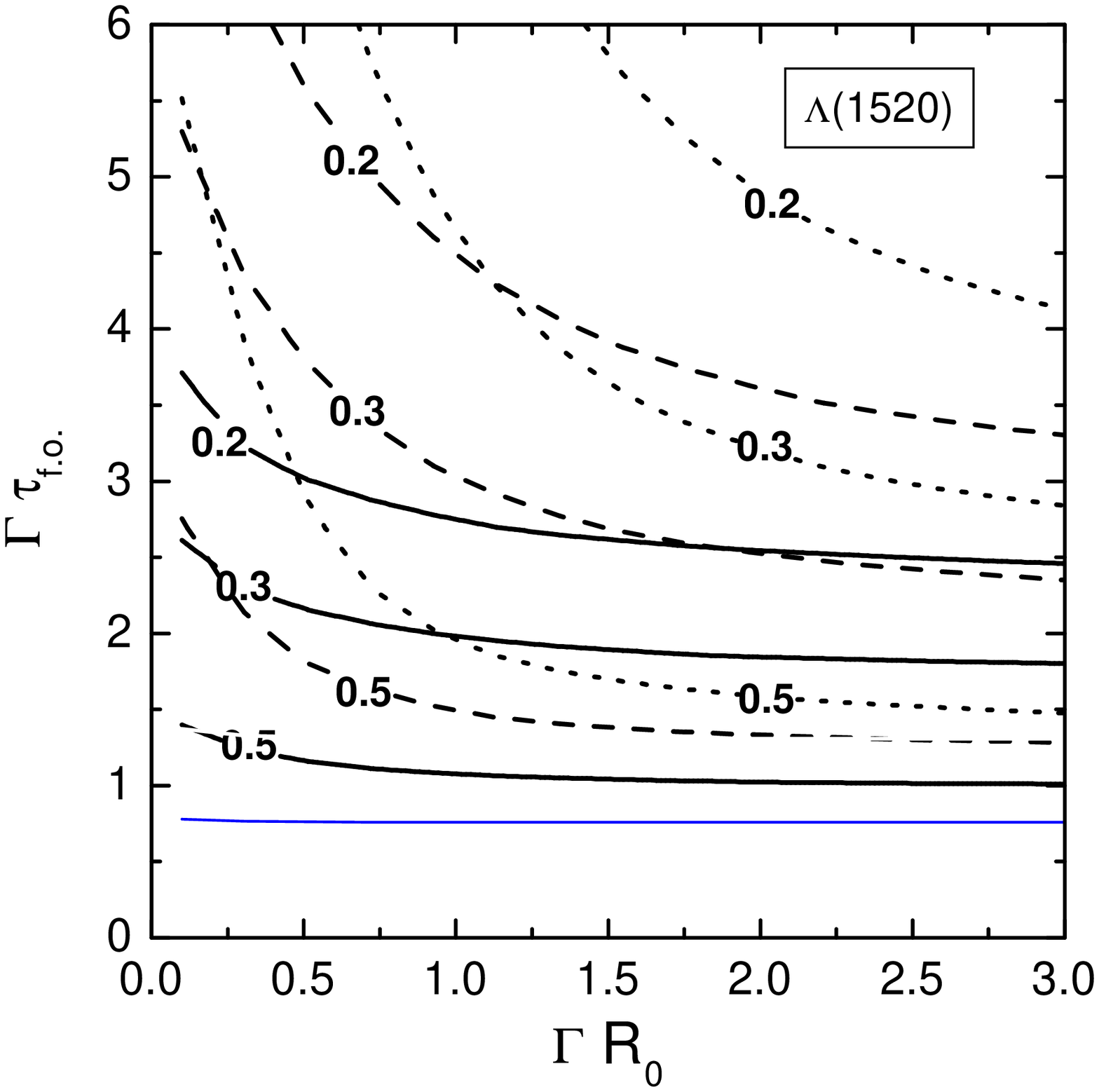}
\includegraphics[width=5.3cm,clip=true]{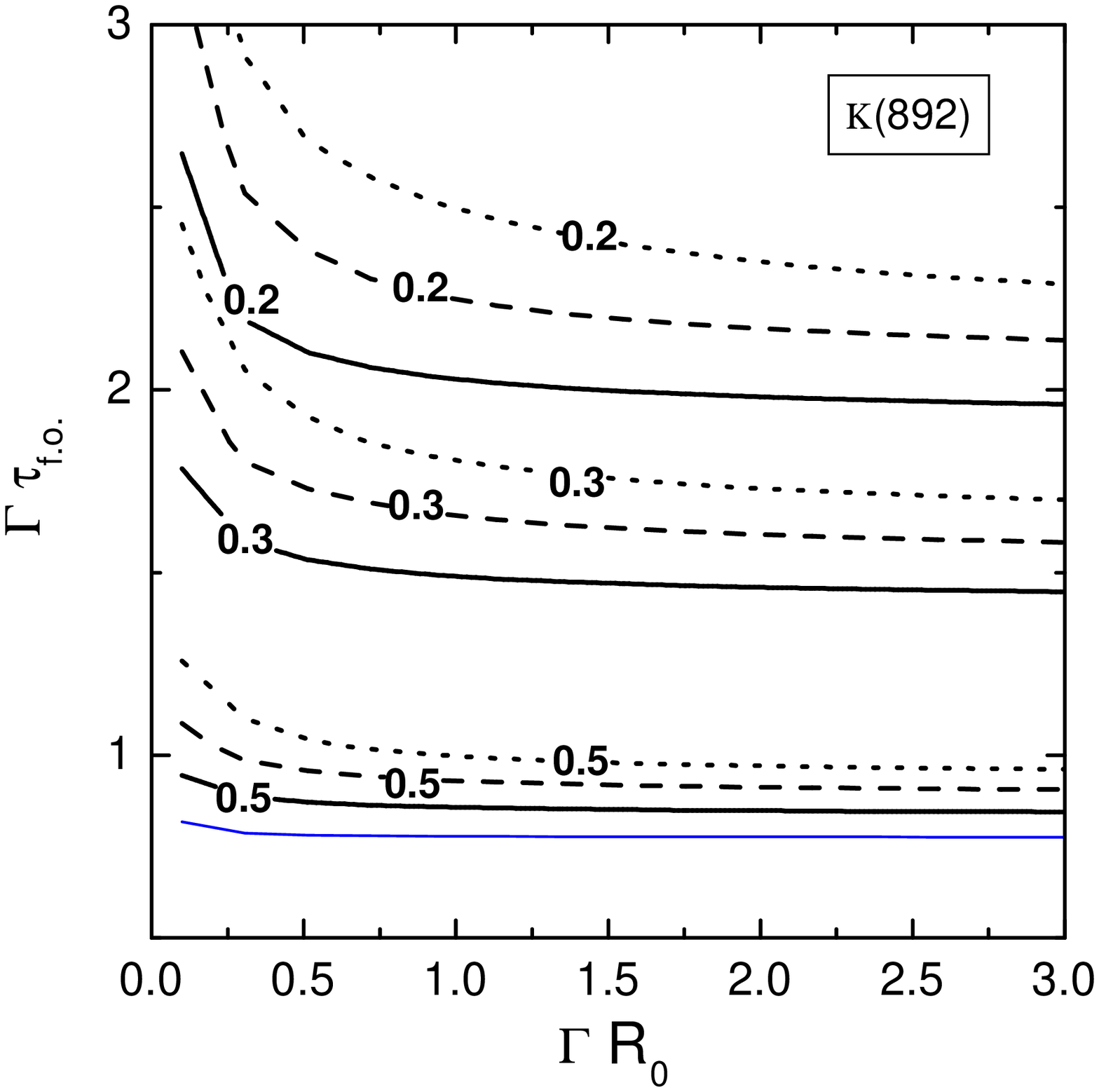}
\caption{Contour plots of ratios $\mathcal{R}^{(r)}(m_r)$ in
(\protect\ref{res}) for  $\phi$ ,
 $\Lambda^*(1520)$ and $\overline{K^*}(892)$ particles on
the plane $\Gamma_{\rm in}^{(r)} R_0$--$\Gamma_{\rm
in}^{(r)}\tau_{\rm f.o.}$. Different line styles correspond to the
different initial resonance widths $\Gamma_{\rm in}$ in (\protect\ref{gam}).
Thick solid, dashed and dotted lines are
drawn, respectively, for $\Gamma_{\rm in}^{(\phi)}=10$, 20,
30~MeV, $\Gamma_{\rm in}^{(\Lambda^*)}=30$, 60, 120~MeV, and
$\Gamma_{\rm in}^{(K^*)}=60$, 70, 80~ MeV.
The thin solid lines are calculated with the vacuum width of the resonance and stand
for the level $\mathcal{R}^{(r)}=0.5$.
All calculations are done for $T=150$~MeV, $v_f=0.5$, and ${\rm P}_{\rm rec}=0$.}
\label{fig:map}
\end{figure}

Let us now consider closer  dependence of the suppression factor on
the fireball  parameters $\tau_{\rm f.o.}$ and $R_0$.
First, we simplify the time dependence of the total width. We use a
linear in density interpolation between the initial value at the
freeze-out moment, $\Gamma_{\rm  in}^{(r)}$ and the vacuum value
\be\label{gam}
\Gamma_{\rm tot}^{(r) *}(t)=
\Gamma_{\rm tot}^{(r)0}+
(\Gamma_{\rm in}^{(r)}-\Gamma_{\rm tot}^{(r) 0})\,(R_0/R(t))^3
\,.
\ee
This approximation works still well and can reproduce the right
panel of Fig.~\ref{fig:phi}, using  $\Gamma_{\rm in}=20$~MeV
within 5--10\% accuracy.
We will vary parameters $R_0$ and $\tau_{\rm f.o.}$ keeping other
parameter fixed: $T_0=150$~MeV, $v_f=0.5$, and  ${\rm P}_{\rm rec}=0$.
In Fig.~\ref{fig:map} we present the contour plot of the value
$\mathcal{R}_K(m_T=m_\phi)$ as a function the dimensionless
variables $\Gamma_{\rm in} \,\tau_{\rm f.o.}$ and $\Gamma_{\rm
in}\, R_0$, calculated for $\Gamma_{\rm in}^{(\phi)}=10$, 20, 30~MeV.
The thin solid line shows the level $\mathcal{R}_K(m_T=m_\phi)=0.5$
for the vacuum width $\Gamma_{\rm in}^{(\phi)}=\Gamma^{(\phi) 0}_{\rm
tot}=4.43$~MeV. We see that for the vacuum width the 50\%
suppression can be reached only for rather large
$\tau_{\rm f.o}\simeq 40$~fm. From this plot
we can also read out the minimal combinations of parameters $(R_0,\tau_{\rm
f.o.})$ required to attain a certain degree of suppression.
For instance, $\mathcal{R}_K(m_\phi)<0.2$ will be reached
(we pick out the points closest to the origin on the lines labeled with 0.2)
for  $\Gamma_{\rm in}^{(\phi)}=10$~MeV if ($\tau_{\rm f.o.}>67$~fm, $R_0>15$~fm),
for  $\Gamma_{\rm in}^{(\phi)}=20$~MeV if ($\tau_{\rm f.o.}>44$~fm,
$R_0>15$~fm), and for  $\Gamma_{\rm in}^{(\phi)}=30$~MeV if ($\tau_{\rm f.o.}>33$~fm,
$R_0>12$~fm). Have we fixed the size of a fireball to be about
20~fm, the level $\mathcal{R}_K(m_\phi)=0.2$ can be arrived at $\tau_{\rm
f.o.}=$63, 39, 26~fm for $\Gamma_{\rm  in}^{(\phi)}=$10, 20 and 30~MeV,
respectively.

We turn now to the $\Lambda^*(1520)$ and $\overline{K^*}(892)$ resonances.
Their mean free paths in the hot
hadronic matter have not been studied so far, unfortunately.
There is also no additional mechanism like OZI for $\phi$ mesons which would
suppress the $\Lambda^*$ and $\overline{K^*}$ interaction with surrounding
pions, kaons, and nucleons. Moreover, $\pi\Lambda^*$ scattering
has  contributions from s-channel hyperon exchange processes,
particularly the reaction $\pi\Lambda^*\to \Sigma^*(1385)\to
\pi\Lambda^*$ should be operative due to the s-wave
$\pi\Lambda^*\Sigma^*$ coupling. Therefore we can expect that the
mean free paths of $\Lambda^*$ and $\overline{K^*}$ are comparable with those
of pions and kaons. Hence their final momentum distribution should
be formed rather close to the common break-up of the fireball.

In Fig.~\ref{fig:map} we depict the contour plots of the quantity
\be\label{res} \mathcal{R}^{(r)}(m_T)=<\eta_{\rm
hadr.}^{(r)}(p)>_y/<\eta^{(r)}_0(p)>_y \ee evaluated for
$m_T=m_r$, $T=150$~MeV, $v_f=0.5$. The apparent distribution in
the dominant hadronic channel $\eta_{\rm hadr.}^{(r)}$ is
calculated according to (\ref{ratrescat})  with ${\rm P}_{\rm
rec}=0$ and  depends only on the  total width of the resonance.
The time dependence of the width is given by (\ref{gam}) with
$\Gamma_{\rm in}^{(r)}$ varying in the broad range separately for
each resonance. The last panel in Fig.~\ref{fig:map} shows us that
to have  for $\overline{K^*}$ mesons  the suppression factor
$\mathcal{R}^{(K^*)}(m_{K^*})>0.5$, we should have  $\Gamma_{\rm
in}^{(K^*)}\, \tau_{\rm in}\sim$0.6--1. This translates for
$\Gamma_{\rm in}^{(K^*)}\sim $50--80 MeV into $\tau_{\rm f.o.}\sim
2$--2.4~fm. Hence the $\overline{K^*}$ mesons should indeed be produced
close to the fireball break up. Using these values for $\Lambda^*$
we find  from the middle plane in Fig.~\ref{fig:map} that the
ratio $\mathcal{R}^{(\Lambda^*)}(m_{\Lambda^*})$ becomes  less
than 0.5 only if $\Gamma_{\rm in}^{(\Lambda^*)}\gsim 120$~MeV.
This is a very large value. Although there are indications
that $\Lambda^*(1520)$ properties suffer a strong
modifications in nuclear medium acquiring the total width of the
order 100~MeV at the normal nuclear matter density~\cite{lk},
calculations have been done so far for cold nuclear matter only.
At high temperatures these effects might be reduced. Alternatively
we can suggest that $\Lambda^*(1520)$ hyperons decouple from the
fireball at some earlier stage than $\overline{K^*}$ mesons. Then we obtain
the following estimates for the time between $\Lambda^*$
freeze-out and the fireball breakup: $\tau_{\rm f.o.}\sim 6$~fm for $\Gamma_{\rm
in}^{(\Lambda^*)}\sim$30~MeV and $\tau_{\rm f.o.}\sim 9$~fm for
the vacuum width.

Finally we do not exclude a possibility that
the formation of $J=\frac32^{(-)}$ baryonic state $\Lambda^*(1520)$
in heavy-ion collisions is already primarily suppressed.
The purpose of the presented work is to find out to what extent
the DPR suppression mechanisms enhanced by the in-medium modification of
resonance width is able to accommodate experimental data.

\section{Summary}

We investigate  dependence of the decay-product rescattering mechanism of
the resonance production  suppression on the size of the hadronic fireball
and on the time between resonance freeze-out and fireball break-up.
Possible modification of the resonance width in medium is shown to
enhance the suppression.
The model is applied to the production of $\phi(1020)$,
$\Lambda^*(1520)$ and $\overline{K^*}(892)$ particles at SPS energies.
We conclude that in the case of $\phi(1020)$ mesons  the discrepancy
between $\phi $ momentum distributions observed in $K^+K^-$ (NA49) and
leptonic (NA50) channels can be explained if at the moment of the $\phi$ freeze-out
the $\phi$ total decay width is about 30~MeV and after freeze-out  the fireball
lives about 20~fm/c till the break-up.
The observed (NA49) signals of  $\overline{K^*}(892)$ meson production indicate
that $\overline{K^*}$ mesons escape from the fireball at the moment very close to the
breakup (not further than 2.0-2.4 fm/c). Otherwise their apparent distribution
would be strongly suppressed.
We can obtain the attenuation of the $\Lambda^*(1520)$ on the level of 50\%
assuming either a very large width of the resonance $\gsim 120$~MeV at the
freeze-out moment close to fireball break-up or that $\Lambda^*$
particles leave the fireball at least 6--9~fm/c before breakup.

We conclude that the resonance production in heavy ion collisions
can serve as an indicator of the fireball dynamics at
the last stage of heavy-ion collision.
To draw precise quantitative conclusions careful investigation
of the modification of resonance properties in hadronic medium is necessary.


\end{document}